# General First-Principles Approach to Crystals in Finite Magnetic Fields

Chengye Lü[1†], Yingwei Chen[1†], Yuzhi Wang[2,3], Zhihao Dai[1], Zhong Fang[2,3], Xin-Gao Gong[1], Quansheng Wu[2,3*], Hongjun Xiang[1*]

*[1]Key Laboratory of Computational Physical Sciences (Ministry of Education), Institute of Computational Physical Sciences, State Key Laboratory of Surface Physics, and Department of Physics, Fudan University, Shanghai, 200433, China*

*[2] Beijing National Laboratory for Condensed Matter Physics and Institute of Physics, Chinese Academy of Sciences, Beijing 100190, China*

*[3] University of Chinese Academy of Sciences, Beijing 100049, China*

*E-mail: quansheng.wu@iphy.ac.cn

*E-mail: hxiang@fudan.edu.cn

† These authors contributed equally to this work.

## ABSTRACT

We introduce a general first-principles methodology for computing electronic structure in a finite uniform magnetic field which allows for an arbitrary rational magnetic flux and nonlocal pseudopotentials, at a comparable time complexity of conventional plane-wave pseudopotential approaches in zero-field conditions. The versatility of this method is demonstrated through comprehensive applications to both molecular and crystalline systems, including calculations of magnetizabilities, magnetically induced currents, and magnetic energy bands. Furthermore, we provide rigorous proofs of two properties for crystals in uniform magnetic fields: the "strong translational symmetry" and "magnetic bands shift" phenomena.

## Introduction

The behavior of matter under external electromagnetic fields has attracted sustained interest. Unlike electric fields, magnetic fields couple independently to the spin and orbital degrees of freedom (DOF) of charged particles. The former, usually termed the spin-Zeeman field, governs conventional magnetism, whereas the latter—referred to here as the magnetic field—underlies phenomena such as the Landau level [1], Abrikosov vortex [2], Quantum Hall Effect (QHE) [3], Fractional QHE [4], strong-field phase transitions [5], and fragile topology [6].

Kohn–Sham Density Functional Theory (KS-DFT) [7,8] is the standard framework for first-principles studies in condensed matter physics and material science. While its extensions to finite electric [9,10] and spin-Zeeman fields [11,12] are well developed, corresponding progress for finite magnetic fields has been limited. London orbital approaches have advanced molecular systems [13–16], but crystalline systems face the complication that magnetic fields break Bloch translational symmetry under periodic boundary conditions (PBC), rendering plane-wave (PW) bases unsuitable. Consequently, most studies employ perturbation theory [17,18] or tight-binding models with Peierls substitution and other similar phase-changing techniques [19–25], which are restricted to weak fields [26–28] and cannot capture nonperturbative phenomena [29,30].

In 2004, Cai *et al.* proposed a novel *ab initio* approach for periodic systems under a finite uniform magnetic field, by introducing a so-called "plane-wave-like" basis set which aligns with the magnetic periodic boundary conditions (MPBC) and the magnetic translational symmetry (MTS), to diagonalize the finite-field Hamiltonian [31,32]. However, their approach has limitations: first, their magnetic flux number is not arbitrary but constrained to being a factor of the number of discrete grid-points along the $y$-direction; second, they cannot treat Hamiltonians with nonlocal pseudopotentials which are however indispensable for treating most elements. To integrate nonlocal pseudopotentials into field-containing computations, Pickard *et al*. proposed the gauge-including projector augmented-wave (GIPAW) method [33,34]. Adopting the symmetric gauge, this method has been proven effective for computations in the perturbative regime of an external magnetic field [33–37], but hasn't been extended to the finite-field regime yet.

In this Letter, we present a general finite-field first-principles approach for periodic systems that accommodates arbitrary rational magnetic fluxes and Hamiltonians with or without nonlocal pseudopotentials, through combining the fast Fourier transform (FFT) and real-space algorithms with the GIPAW method under the Landau-type gauge. We present applications of this novel method to both molecular and crystal systems by computing magnetizabilities, magnetically induced currents and magnetic energy bands. The whole algorithm is implemented in a developer's version of the Property Analysis and Simulation Package for materials (PASP) [38].

## Methods

**Hamiltonian and wavefunction-**We adopt the atomic unit system ($\hbar = e = m_e = 1$) throughout this Letter, where $\hbar$ is the reduced Planck constant, $e$ is the elementary charge and $m_e$ is the electron mass. Consider a spin-less crystalline system with lattice periodicities $\boldsymbol{a}_\alpha$ ($\alpha = 1,2,3$), the KS Hamiltonian for the system, which employs Kleinman-Bylander pseudopotentials [39], in the presence of a uniform magnetic field $\boldsymbol{B}$ is written as

$$\hat{H} \coloneqq \frac{1}{2}\hat{\boldsymbol{\pi}}^2 + V_{\text{per}}(\boldsymbol{r}) + \hat{V}_{\text{GIPAW}}, \tag{1}$$

where $\hat{\boldsymbol{\pi}} \coloneqq -\text{i}\boldsymbol{\nabla} + \boldsymbol{A}(\boldsymbol{r})$ is the mechanical momentum operator and $\boldsymbol{A}(\boldsymbol{r})$ is the magnetic vector potential satisfying the condition $\nabla \times \boldsymbol{A} = \boldsymbol{B}$ and the linearity $\boldsymbol{A}(\boldsymbol{r}_1 + \boldsymbol{r}_2) = \boldsymbol{A}(\boldsymbol{r}_1) + \boldsymbol{A}(\boldsymbol{r}_2)$. The periodic local potential $V_{\text{per}}(\boldsymbol{r})$, defined by $V_{\text{per}}(\boldsymbol{r}) = V_{\text{per}}(\boldsymbol{r} + \boldsymbol{a}_\alpha)$ with $\boldsymbol{a}$ as the lattice vector, consists of three terms: the local pseudopotential, the Hartree potential and the exchange-correlation potential. Here we neglect the dependence of exchange-correlation functional on the magnetic-field or current, and a detailed discussion of density functional approximations in a magnetic field is beyond the scope of this work [40–45]. The final term $\hat{V}_{\text{GIPAW}}$ (to be discussed below) represents the non-local contribution from the GIPAW method which ensures the recovery of MTS.

As discovered in previous studies, when the magnetic field satisfies the magnetic flux quantization condition, i.e., $\boldsymbol{B}\cdot(\boldsymbol{a}_1\times\boldsymbol{a}_2)=2\pi n_\Phi$ where $n_\Phi\in\mathbb{Z}$ is the magnetic flux number, the eigenstates $\{\psi_{i\boldsymbol{k}}\}$ of the Hamiltonian, corresponding to the eigenvalues $\{\epsilon_{i\boldsymbol{k}}\}$, obey the magnetic Bloch theorem (MBT) [31,32,46–50]

$$\psi_{i\boldsymbol{k}}(\boldsymbol{r})=\mathrm{e}^{\mathrm{i}\boldsymbol{k}\cdot\boldsymbol{r}}u_{i\boldsymbol{k}}(\boldsymbol{r}), \tag{2.1}$$

$$u_{i\boldsymbol{k}}(\boldsymbol{r}+\boldsymbol{a}_\alpha)=\mathrm{e}^{-\mathrm{i}\boldsymbol{A}(\boldsymbol{a}_\alpha)\cdot\boldsymbol{r}}u_{i\boldsymbol{k}}(\boldsymbol{r}), \tag{2.2}$$

where $i$ denotes the band index and $\boldsymbol{k}$ belongs to the first Magnetic Brillouin Zone (FMBZ). Our task is to solve the KS Hamiltonian (1) under the MPBC described by equations (2.2) through the self-consistency-field (SCF) loop. It is important to note that the theorem can be generalized to any rational values of $n_\Phi=p/q$ with $p$ and $q$ coprime since it's always possible to enlarge the unit cell by a factor of $q$ to construct a magnetic unit cell that quantizes the magnetic flux. The corresponding FMBZ of the magnetic unit cell is also $q$ times smaller than the first Brillouin Zone (FBZ) of the chemical unit cell. Moreover, it's also helpful to note that the electronic density $\rho(\boldsymbol{r})$ and the physical current density $\boldsymbol{j}(\boldsymbol{r})$ generated by magnetic Bloch functions are periodic with respect to the magnetic unit cell:

$$\rho(\boldsymbol{r})\coloneqq\sum_{i\boldsymbol{k}}f_{i\boldsymbol{k}}|\psi_{i\boldsymbol{k}}(\boldsymbol{r})|^2=\rho(\boldsymbol{r}+\boldsymbol{a}_\alpha), \tag{3.1}$$

$$\boldsymbol{j}(\boldsymbol{r})\coloneqq-\sum_{i\boldsymbol{k}}f_{i\boldsymbol{k}}\Re[\psi_{i\boldsymbol{k}}^*(\boldsymbol{r})\hat{\boldsymbol{\pi}}\psi_{i\boldsymbol{k}}(\boldsymbol{r})]=\boldsymbol{j}(\boldsymbol{r}+\boldsymbol{a}_\alpha), \tag{3.2}$$

where $f_{i\boldsymbol{k}}$ denotes the occupation number and $\Re$ represents the real part. Remarkably, one can prove that both $\rho(\boldsymbol{r})$ and $\boldsymbol{j}(\boldsymbol{r})$ are periodic not only within the magnetic unit cell but also within the chemical unit cell. This theorem, which we refer to as "strong translational symmetry", is proven in the supplementary materials (SM) [51].

In our approach, we assume that the magnetic field is oriented along the $\boldsymbol{a}_3$ direction, i.e., $\boldsymbol{B}=B\hat{\boldsymbol{a}}_3=2\pi n_\Phi\boldsymbol{a}_3/\Omega$, and we adopt the Landau-type gauge [32,52,53] defined by $\boldsymbol{A}_{\mathrm{Lan}}(\boldsymbol{r})\coloneqq 2\pi n_\Phi(\boldsymbol{b}^1\cdot\boldsymbol{r})\boldsymbol{b}^2$, where $\hat{\boldsymbol{a}}_3$ is the unit vector of $\boldsymbol{a}_3$-direction, $\Omega\coloneqq\boldsymbol{a}_3\cdot(\boldsymbol{a}_1\times\boldsymbol{a}_2)$ is the volume of the magnetic unit cell and $\boldsymbol{b}^\alpha$ $(\alpha=1,2,3)$ are the (scaled) reciprocal basis vectors defined by $\boldsymbol{b}^\alpha\cdot\boldsymbol{a}_\beta=\delta_\beta^\alpha$ for $\alpha,\beta=1,2,3$. Under this gauge, the MPBC is explicitly written as

$$\begin{gathered}u_{i\boldsymbol{k}}(\xi^1+1,\xi^2,\xi^3)=\mathrm{e}^{-\mathrm{i}2\pi n_\Phi\xi^2}u_{i\boldsymbol{k}}(\xi^1,\xi^2,\xi^3),\\ u_{i\boldsymbol{k}}(\xi^1,\xi^2+1,\xi^3)=u_{i\boldsymbol{k}}(\xi^1,\xi^2,\xi^3+1)=u_{i\boldsymbol{k}}(\xi^1,\xi^2,\xi^3),\end{gathered} \tag{4}$$

where $\boldsymbol{r} = \sum_{\alpha=1}^{3} \xi^\alpha \boldsymbol{a}_\alpha$, with $\xi^\alpha \in [0,1)$. Our method can, in principle, be generalized to any gauge. Since the local potential in Eq. (1) acts as a direct multiplication in real-space, we focus on the algorithms for the kinetic and nonlocal terms in the following paragraphs.

**Algorithm on kinetic operator-** When a magnetic field is applied, the kinetic term involves the vector potential, and the MPBC in Eq. (4) introduces non-periodicity along the $\boldsymbol{a}_1$ direction. Interestingly, similar FFT-based techniques can still be utilized by introducing a set of auxiliary functions [54] defined as $h_{i\boldsymbol{k}}^{n_\Phi}(\xi^1,\xi^2,\xi^3) \coloneqq \exp(\mathrm{i}2\pi n_\Phi \xi^1 \xi^2)\, u_{i\boldsymbol{k}}(\xi^1,\xi^2,\xi^3)$, which are periodic along $\boldsymbol{a}_1$ and $\boldsymbol{a}_3$ but non-periodic along $\boldsymbol{a}_2$. To formulate the algorithm, we rewrite the kinetic operator as $\frac{1}{2}\sum_{\alpha\beta} g^{\alpha\beta} \hat{\pi}_\alpha \hat{\pi}_\beta$, where $\hat{\pi}_\alpha \coloneqq \boldsymbol{a}_\alpha \cdot \hat{\boldsymbol{\pi}} = -\mathrm{i}\partial/\partial\xi^\alpha + A_\alpha$ and $A_\alpha \coloneqq \boldsymbol{a}_\alpha \cdot \boldsymbol{A}$ represent the covariant components of the kinetic momentum and vector potential respectively, and $g^{\alpha\beta} \coloneqq \boldsymbol{b}^\alpha \cdot \boldsymbol{b}^\beta$ represents the contravariant metric tensor. To be more specific: for $\hat{\pi}_2$ and $\hat{\pi}_3$, the derivatives $\partial/\partial\xi^\alpha$ and vector potentials $A_\alpha$ are computed through standard FFT and direct multiplication in real-space, respectively; for $\hat{\pi}_1 = -\mathrm{i}\partial/\partial\xi^1$, we use the relation that

$$\hat{\pi}_1 u_{i\boldsymbol{k}} = -2\pi n_\Phi \xi^2 u_{i\boldsymbol{k}} + \mathrm{e}^{-\mathrm{i}2\pi n_\Phi \xi^1 \xi^2} \left(-\mathrm{i}\frac{\partial}{\partial \xi^1} h_{i\boldsymbol{k}}^{n_\Phi}\right), \tag{5}$$

where the first term can be computed via real-space multiplication and the second term via FFT due to the $\xi^1$-periodicity of $h_{i\boldsymbol{k}}^{n_\Phi}$. Thus, the entire kinetic term can be efficiently computed by applying these methods step by step (see Fig. 1).

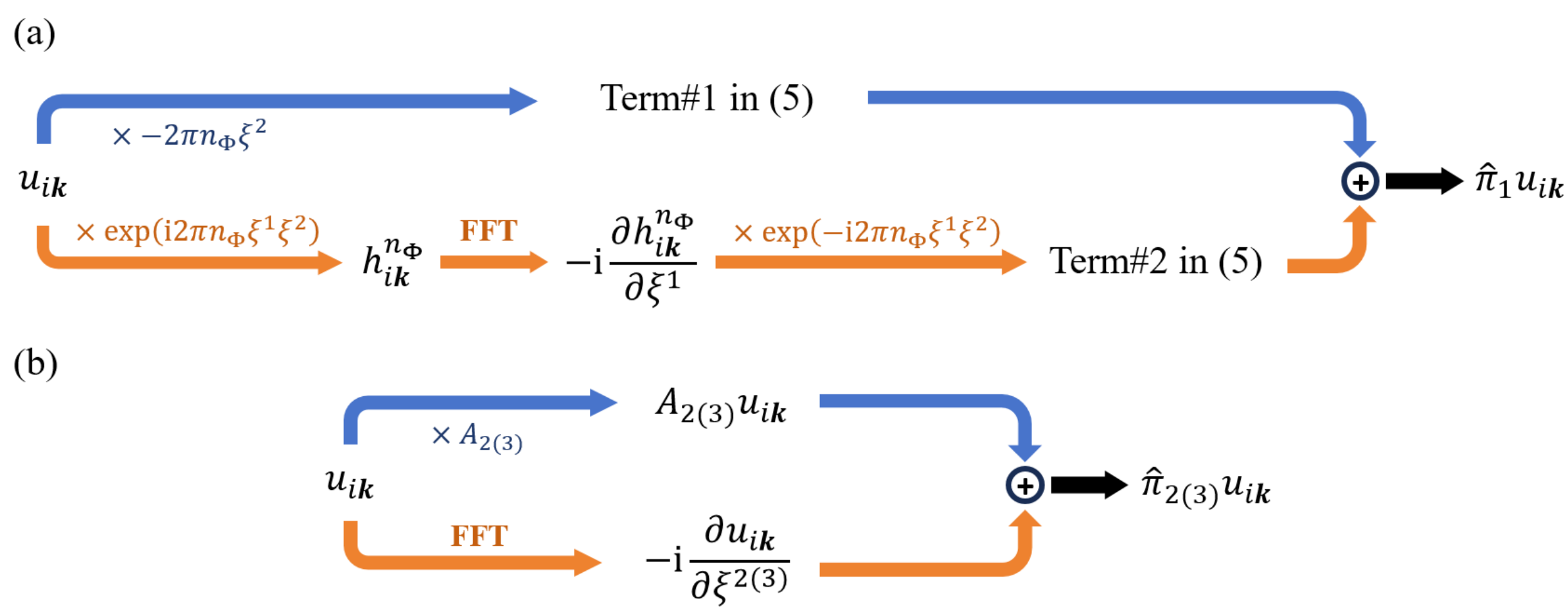

FIG.1. Flowchart for the algorithm on applying kinetic momentum operators to the wavefunction. (1) algorithm for $\hat{\pi}_1$. (2) algorithm for $\hat{\pi}_2$ and $\hat{\pi}_3$.

**Algorithm on nonlocal pseudopotentials-**The nonlocal term by the GIPAW formulation writes as $\hat{V}_{\mathrm{GIPAW}} \coloneqq \sum_{\boldsymbol{R},n} |\tilde{p}_{\boldsymbol{R},n}\rangle a_n^{\boldsymbol{R}} \langle \tilde{p}_{\boldsymbol{R},n}|$ with $\{a_n^{\boldsymbol{R}}\}$ the nonlocal strength on each atomic site $\boldsymbol{R}$ and each channel $n$, and the corresponding nonlocal projectors $\{\tilde{p}_{\boldsymbol{R},n}\}$ defined as

$$\tilde{p}_{\boldsymbol{R},n}(\boldsymbol{r}) \coloneqq \exp[-\mathrm{i}\boldsymbol{A}_{\mathrm{Lan}}(\boldsymbol{R}) \cdot \boldsymbol{r} - \mathrm{i}f(\boldsymbol{r}-\boldsymbol{R})]\, p_{\boldsymbol{R},n}(\boldsymbol{r}), \tag{6}$$

where $\{p_{\boldsymbol{R},n}\}$ represents the nonlocal pseudopotential projectors without magnetic fields. The first phase factor, $\exp[-\mathrm{i}\boldsymbol{A}_{\mathrm{Lan}}(\boldsymbol{R}) \cdot \boldsymbol{r}]$, in Eq. (6) properly restores the MTS of nonlocal pseudopotentials under the magnetic field [33,34]. The second phase factor, $\exp[-\mathrm{i}f(\boldsymbol{r}-\boldsymbol{R})]$, where $\nabla f(\boldsymbol{r}) \coloneqq \boldsymbol{A}_{\mathrm{Lan}}(\boldsymbol{r}) - \boldsymbol{A}_{\mathrm{sym}}(\boldsymbol{r})$ denotes the gauge transformation between the symmetric and Landau-type gauge, introduces an extra phase when using a non-symmetric gauge, such as the Landau-type gauge in our approach. Inclusion of this second phase factor is essential because field-free orbitals (or projectors) can only approximate magnetic orbitals (or projectors) under the symmetric gauge $\boldsymbol{A}_{\mathrm{sym}}(\boldsymbol{r}) \coloneqq \boldsymbol{B} \times \boldsymbol{r}/2$ owing to symmetry considerations, as noted in Ref. [48]. Moreover, since we only employ norm-conserving pseudopotentials [55] in this work, the GIPAW correction to the non-local strength $a_n^{\boldsymbol{R}}$ can be safely omitted [33]. We follow the conventional real-space implementation of nonlocal pseudopotentials [56,57] in our approach, with the key distinction that nonlocal projectors incorporate additional phases as defined in Eq. (6).

To sum up, the local potential and GIPAW-type nonlocal pseudopotential operate in the real space while the kinetic operator operates via both the real-space multiplication and the FFT technique with the help of auxiliary functions.

## Results

**Test on molecules-**We begin by validating our method via calculating the magnetizabilities for both closed-shell diamagnetic and closed-shell paramagnetic molecules. Magnetizability can be evaluated using the second-order finite-difference method within our finite-field framework

$$\chi_B \coloneqq -\frac{\partial^2 E(B)}{\partial B^2}\Big|_{B=0} \approx -\frac{E(\Delta B) - 2E(0) + E(-\Delta B)}{\Delta B^2}, \tag{7}$$

where $\Delta B$ denotes the magnetic field in the supercell at $n_\Phi = 1$ ($\approx 2871$ T). We ensure $E(\Delta B)$ equals to $E(-\Delta B)$ within numerical accuracy in our computations. Note that, in principle, only the magnetizability contributed by valence electrons can be computed via the pseudopotential approximation, while for elements lighter than neon the core contributions can be neglected [17]. The magnetizabilities, computed via PBE functional [58], are listed in the Table 1. All details of the calculations in this work can be found in SM [51]. The molecular configurations and comparative results at finite-field HF and Coupled-Cluster-Single-Double (CCSD) levels using London orbitals are extracted from Ref. [43]. Our approach gives comparable results to all molecules we consider.

To further validate our finite-field DFT method in the molecular level, we calculate the magnetically induced current density of benzene perpendicular to the magnetic field, known as ring currents [59], as plotted in Fig. 2(a). We observe a primary diamagnetic vortex flowing around the molecule's tails, small diamagnetic vortexes surrounding the C-H and C-C bonds and a paramagnetic vortex rotating in the opposite direction at the center. Our results generally agree well with those in Ref. [60] which used London orbitals. The discrepancy near the carbon nuclei should be attributed to our neglect of core electrons. We also discuss the variation of benzene's electron density under the field in SM [51].

| | PBE-DFT | HF | CCSD |
|---|---|---|---|
| Be | -227.0 | -227.8 | -214.0 |
| $H_2$ | -71.2 | -69.2 | -68.4 |
| CO | -299.2 | −299.8 | −300.0 |
| HF | -175.4 | -175.5 | -178.5 |
| $H_2O$ | -239.5 | −232.9 | −235.4 |
| AlH | 269.5 | 277.3 | 233.5 |
| BH | 504.6 | 560.2 | 411.2 |

Table 1. Molecular magnetizabilities calculated at the PBE-DFT, HF and CCSD levels of theory in units of $10^{-30}$ J $T^{-2}$. PBE-DFT results are obtained by our finite-field approach while HF and CCSD results are extracted from Ref. [43] in which London orbitals are used.

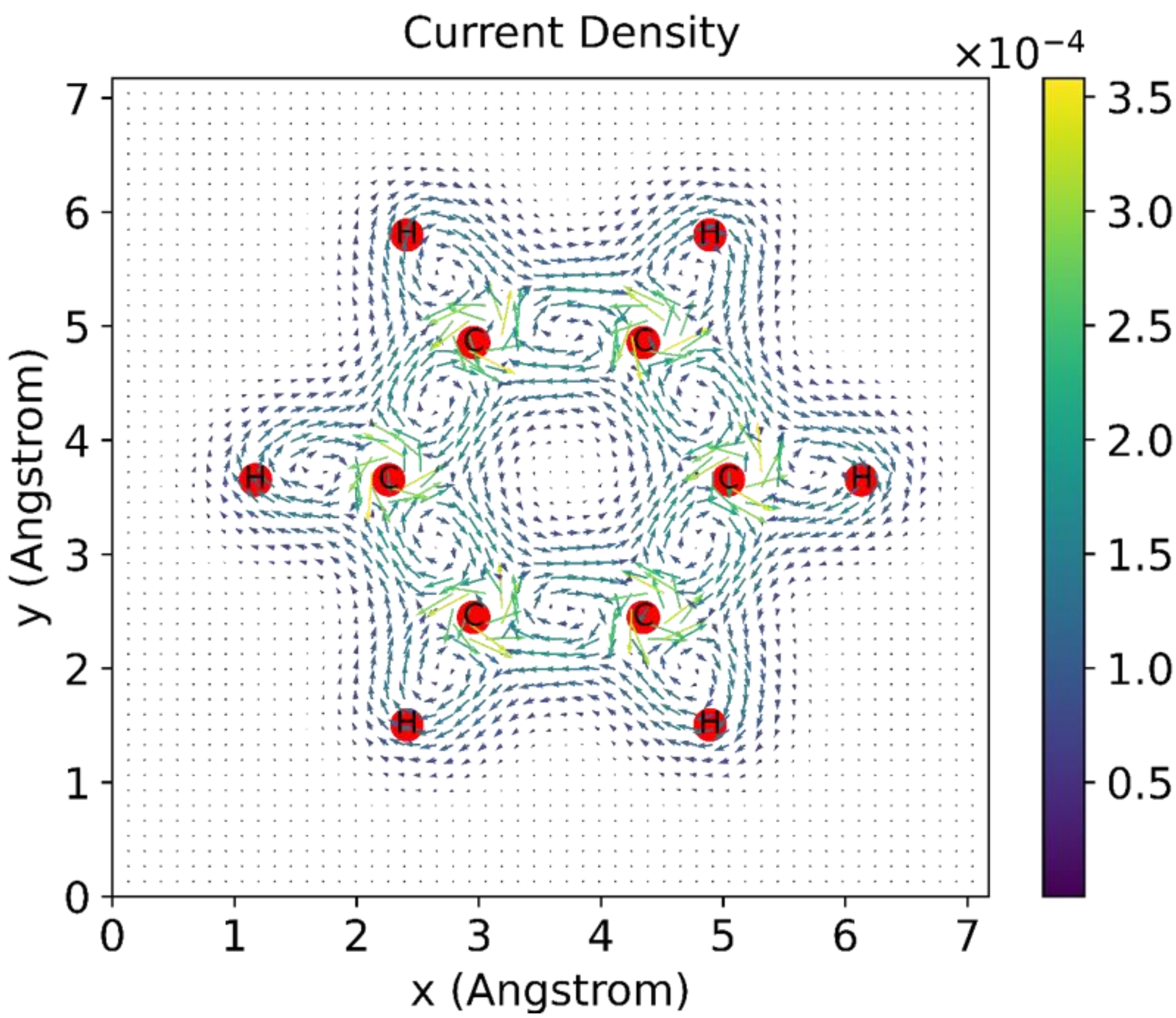

FIG. 2. Magnetically induced valence current density of benzene in a uniform magnetic field of approximately 2871 T.

**Test on solids**-Similar to the molecular test, we first compute the (valence) magnetizabilities for solids including inert-gas crystals and semiconductors, with the results presented in Table 2. The core contribution to the magnetizabilities, valence contribution computed via all-electron perturbation theory within the local density approximation (LDA) [61], and experimental data are all extracted from Ref. [17]. Our finite-field approach yields magnetizabilities for argon and krypton comparable to those via the all-electron perturbative method. For diamond and silicon, our finite field results agree better with the experimental values than the previous results from the perturbative method. In particular, $\chi^{\text{finite field}}$, $\chi^{\text{perturb}}$, $\chi^{\text{expt}}$ for silicon are -5.0, -1.2, -6.4 (in units of $10^{-6}\ \text{cm}^3/\text{mole}$), respectively. The finite-field approach for computing silicon magnetizability outperforms the perturbative method by requiring only total energies, without relying on the band gap which is often poorly captured in DFT calculations [17].

| | Lattice constant (Å) | $\chi_{\text{core}}$ | $\chi_{\text{val}}^{\text{finite field}}$ | $\chi_{\text{val}}^{\text{perturb}}$ | $\chi^{\text{finite field}}$ | $\chi^{\text{perturb}}$ | $\chi^{\text{expt}}$ |
|---|---|---|---|---|---|---|---|
| Ar | 5.36 | -1.19 | -19.2 | -19.97 | -20.39 | -20.84 | - |
| Kr | 5.79 | -5.41 | -24.6 | -24.90 | -30.01 | -30.31 | - |
| diamond | 3.56 | -0.3 | -11.2 | -10.3 | -11.5 | -10.6 | -11.8 |
| silicon | 5.44 | -4.8 | -0.2 | 3.6 | -5.0 | -1.2 | -6.4 |

Table 2. Solid-state magnetizabilities computed via finite-field and perturbative approaches within LDA-DFT in units of $10^{-6}\,\text{cm}^3/\text{mole}$. The core and valence contributions obtained via all-electron perturbation theory, along with the experimental results, are extracted from Ref. [17].

Subsequently, we compute the magnetic band structure of graphene to illustrate the impact of a finite field. Figs. 3(a) and 3(b) show the bands obtained without and with a magnetic field of $n_\Phi = 1$ (approximately 80125 T), using non-SCF calculations based on field-free and field-containing SCF charge densities, respectively. The pronounced flattening of the bands under the field is primarily attributed to modifications of the Hamiltonian and boundary conditions induced by the flux, rather than to changes in the density, which remain relatively negligible even at this strength. This is confirmed by Fig. 3(c), where the same field is applied but the density is fixed to the field-free case: the resulting bands and total energies differ only slightly from those of Fig. 3(b), supporting that the effects are flux-driven.

Notably, however, we find that the magnetic bands are not invariant under collective atomic translations. As shown in Fig. 3(d), a translation of $0.5\,\boldsymbol{a}_1$ produces a band structure $\{\epsilon_{i\boldsymbol{k}}|\boldsymbol{k} \in \text{FMBZ}\}$ markedly different from Fig. 3(b), indicating that the bands lack translational invariance in contrast to the field-free scenario. This phenomenon, referred to as "magnetic band shift" in this Letter, arises because the atomic translation modifies only the scalar potential, which is equivalent to a linear gauge transformation of the vector potential, introducing an additional position-dependent phase in the eigenstates. The phase can then be absorbed into a redefinition of $\boldsymbol{k}$-points within the FMBZ, effectively shifting the energy levels onto new pseudo-momenta. Consequently, the absolute shape of the magnetic bands loses physical significance, and its interpretation becomes ambiguous. Nevertheless, the density of states (DOS) as well as the total energy remains translationally invariant, as illustrated in Figs. 3(e) and 3(f), strengthening our argument. Details and additional examples of "magnetic band shift" are provided in SM [51].

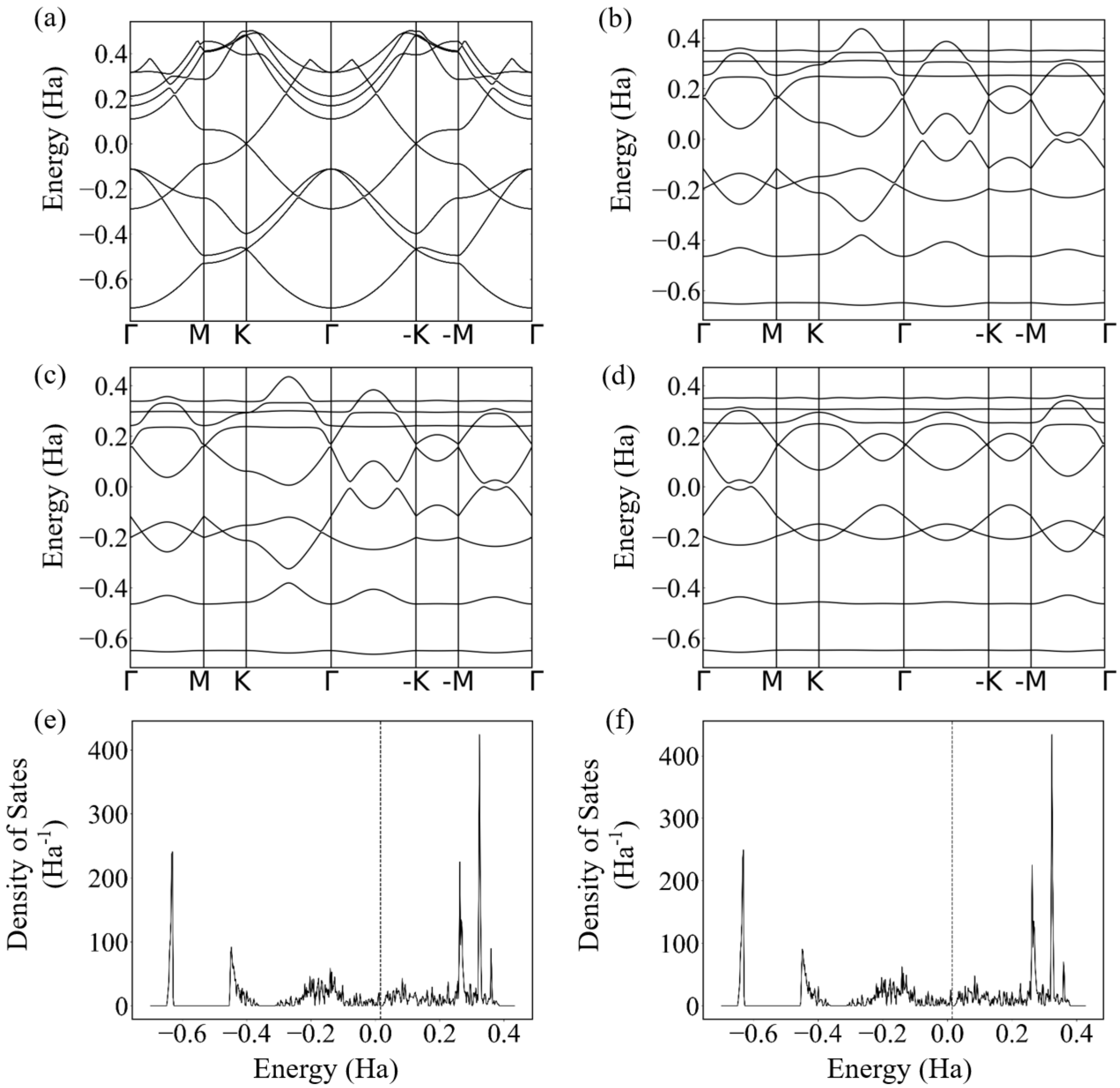

FIG. 3. Magnetic field effects on the band structure of graphene. (a) Electronic band structure in the absence of an external magnetic field. (b) Magnetic band structure under an external magnetic field of $n_\Phi = 1$ (approximately $80125$ T). (c) Magnetic band structure under the same field strength as in (b), but using the electron density from the field-free SCF calculation shown in (a). (d) Magnetic band structure under the same field strength as in (b), following a collective atomic translation of $0.5\,\boldsymbol{a}_1$. The energy levels in (a)-(d) are renormalized to the corresponding Fermi energies. (e), (f) Density of states (DOS) of graphene under the same magnetic field as in (b),

before and after the collective atomic translation of $0.5\,\boldsymbol{a}_1$, respectively. The dotted line indicates the Fermi energy.

Finally, we compute the Berry curvatures and Chern numbers of graphene under a magnetic field of $n_\Phi = 1$, using Fukui's algorithm [62,63]. Although the topologically non-trivial nature of magnetic bands, characterized by non-zero Chern numbers, has been established through earlier mathematical analyses [3,64], it has not yet been verified from a first-principles perspective. As a preliminary validation, we disable all scalar potentials in the Hamiltonian to solve the famous Landau problem and, as a result, successfully recover the equally spaced flat bands with a constant Berry curvature of magnitude $1/B$, whose integration over the FMBZ yields a Chern number of 1 [65]. We then apply this procedure to graphene, whose lowest valence band exhibits a non-constant Berry curvature but still yields a Chern number of 1, thereby confirming the non-trivial topology of magnetic bands. Moreover, the second lowest band yields a Chern number of 0 while the entangled third and fourth lowest bands yield a Chern number of 3 jointly, indicating magnetic bands cannot be naively interpretated as Landau Levels in general. For illustration, the Berry curvatures of the two lowest bands of graphene are presented in Fig. 4.

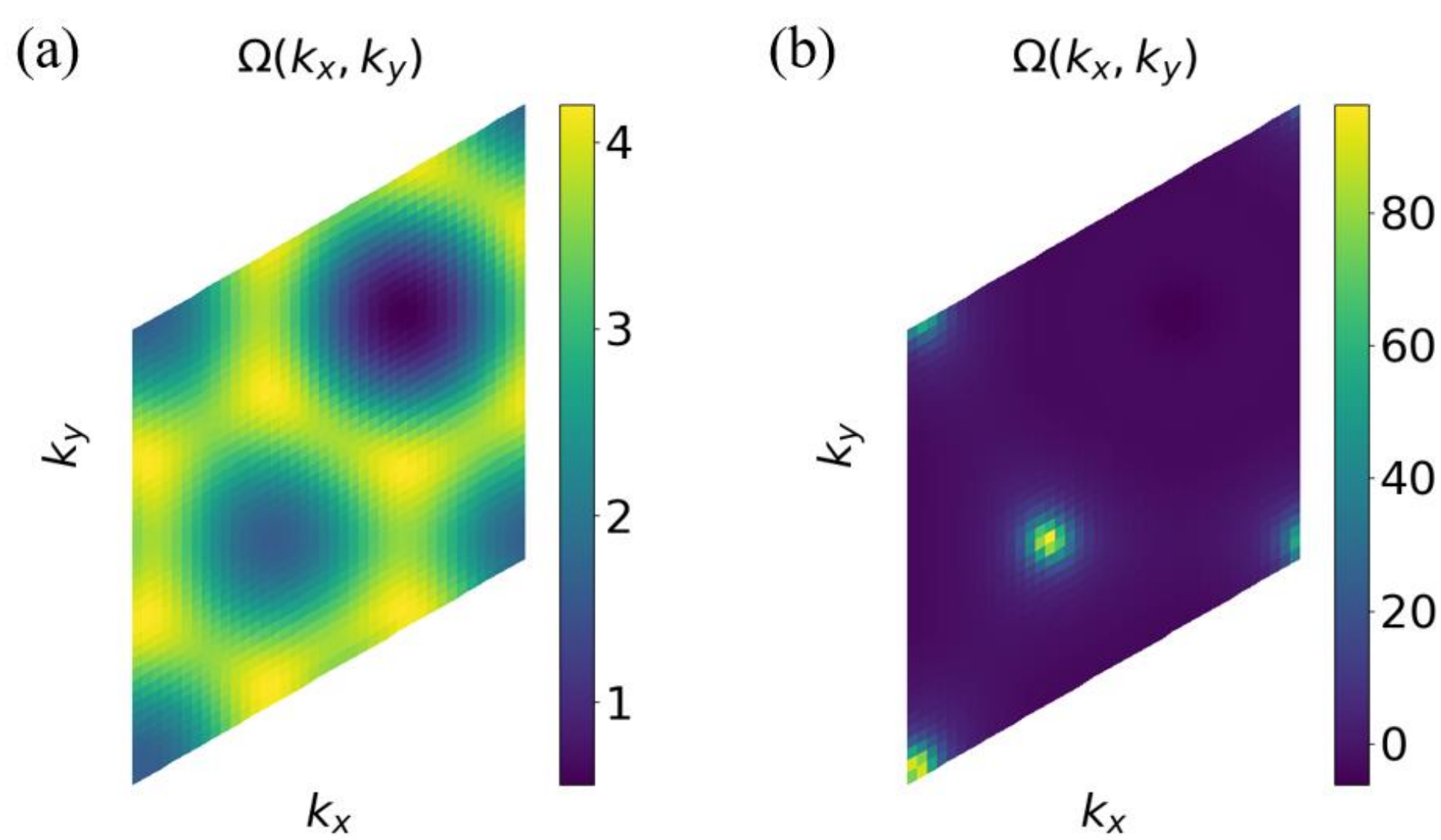

FIG. 4. Berry curvature distributions in graphene under a magnetic field of $n_\Phi = 1$. (a) Berry curvature of the lowest energy band, which carries a Chern number of 1. (b) Berry curvature of the second lowest energy band, which has a Chern number of 0.

## Conclusion

In summary, we have presented a new first-principles scheme for computing the electronic structure of crystals in uniform magnetic fields of an arbitrary rational quantum flux. The algorithm combines FFT technique and real-space methods, at a computational time complexity to that of conventional field-free DFT based on PW basis and pseudopotential approximations. This algorithm is also the first application of the GIPAW method in finite-field regimes. In this work, we consider small magnetic unit cells to reduce computational demanding, and thus the applied magnetic field is rather high. This constraint is expected to be overcome in the near future with advances in large-scale numeric algorithms [66], modern parallel design [67] and machine learning methods [68,69]. The computation of ionic forces, as well as extensions of this method to incorporate other variables like spin-density, current density and spin-current density [70], is straightforward and will also be investigated in our future works.

## Acknowledgement

We acknowledge financial support from NSFC (No. 12188101), the National Key R&D Program of China (No. 2022YFA1402901, 2023YFA1607400), the Guangdong Major Project of the Basic and Applied Basic Research (Future functional materials under extreme conditions—2021B0301030005), Shanghai Science and Technology Program (No. 23JC1400900), and Shanghai Pilot Program for Basic Research—FuDan University 21TQ1400100 (23TQ017), New Cornerstone Science Foundation. We thank Prof. Wei Hu for discussions about DFT and thank Tao Zhang for discussions about quantum mechanics.

## Reference for the Supplemental Material

# Supplementary Material for

## General First-Principles Approach to Crystals in Finite Magnetic Fields

Chengye Lü[1†], Yingwei Chen[1†], Yuzhi Wang[2,3], Zhihao Dai[1], Zhong Fang[2,3], Xin-Gao Gong[1], Quansheng Wu[2,3*], Hongjun Xiang[1*]

[1]*Key Laboratory of Computational Physical Sciences (Ministry of Education), Institute of Computational Physical Sciences, State Key Laboratory of Surface Physics, and Department of Physics, Fudan University, Shanghai, 200433, China*

[2] *Beijing National Laboratory for Condensed Matter Physics and Institute of Physics, Chinese Academy of Sciences, Beijing 100190, China*

[3] *University of Chinese Academy of Sciences, Beijing 100049, China*

*E-mail: quansheng.wu@iphy.ac.cn

*E-mail: hxiang@fudan.edu.cn

† These authors contributed equally to this work.

## 1. Strong translational symmetry for crystal under magnetic field

In this sector, we would prove that the electronic density $\rho(\boldsymbol{r})=\sum_i^{\text{occ}}\sum_{\boldsymbol{k}\in\text{FBZ}} f_{i\boldsymbol{k}}|\psi_{i\boldsymbol{k}}(\boldsymbol{r})|^2$ and the current density $\boldsymbol{j}(\boldsymbol{r})=-\sum_i^{\text{occ}}\sum_{\boldsymbol{k}\in\text{FBZ}} f_{i\boldsymbol{k}}\Re\left[\psi_{i\boldsymbol{k}}^*(\boldsymbol{r})\left(-\mathrm{i}\boldsymbol{\nabla}+\boldsymbol{A}(\boldsymbol{r})\right)\psi_{i\boldsymbol{k}}^*(\boldsymbol{r})\right]$ (which are defined in main text as Eq.(3.1) and Eq.(3.2)) have the periodicity within the chemical unit cell, referred to as "strong translational symmetry" by us. The numerical evidence is also plotted in Fig. S1.

For simplicity, we only consider a two-dimensional system with a rectangular chemical unit cell of size $a_1\times a_2=1\times 1$ and an external magnetic field $B=\frac{2\pi}{a_1a_2}\frac{p}{q}=2\pi\frac{p}{q}$ applied perpendicular to it. The corresponding Landau gauge is $\boldsymbol{A}(\boldsymbol{r})=2\pi\frac{p}{q}x\hat{y}$.

We follow the derivation in Ref. [1] by defining two magnetic translational operators $\widehat{M}(\boldsymbol{a}_1)=\mathrm{e}^{-\mathrm{i}\boldsymbol{A}(\boldsymbol{a}_1)\cdot\boldsymbol{r}}\mathrm{e}^{-\mathrm{i}\boldsymbol{a}_1\cdot\widehat{\boldsymbol{p}}}=\mathrm{e}^{-\mathrm{i}2\pi\frac{p}{q}y}\mathrm{e}^{-\frac{\partial}{\partial x}}$ and $\widehat{M}(\boldsymbol{a}_2)=\mathrm{e}^{-\mathrm{i}\boldsymbol{A}(\boldsymbol{a}_2)\cdot\boldsymbol{r}}\mathrm{e}^{-\mathrm{i}\boldsymbol{a}_2\cdot\widehat{\boldsymbol{p}}}=\mathrm{e}^{-\frac{\partial}{\partial y}}$. The two operators both commute with Hamiltonian $\widehat{H}$ but don't commute with each other in general. In fact, $\widehat{M}(\boldsymbol{a}_1)\widehat{M}(\boldsymbol{a}_2)=\mathrm{e}^{-\mathrm{i}2\pi\frac{p}{q}}\widehat{M}(\boldsymbol{a}_2)\widehat{M}(\boldsymbol{a}_1)$. Nevertheless, we can enlarge the chemical unit cell along $x$-direction by a factor of $q$ to construct the magnetic unit cell, and in this circumstance $\widehat{M}(q\boldsymbol{a}_1)\left(=\widehat{M}^q(\boldsymbol{a}_1)\right)$ commutes with $\widehat{M}(\boldsymbol{a}_2)$ (it can be proved that $\widehat{M}(q\boldsymbol{a}_2)$ commutes with $\widehat{M}(\boldsymbol{a}_1)$ when expanding the cell along another direction, too). Therefore, we can find common eigenstates $\psi_{i,k_1,k_2}$ for $\{\widehat{M}(q\boldsymbol{a}_1), \widehat{M}(\boldsymbol{a}_2), \widehat{H}\}$ with corresponding eigenvalues $\{\mathrm{e}^{-\mathrm{i}qk_1}, \mathrm{e}^{-\mathrm{i}k_2}, \epsilon_{i\boldsymbol{k}}\}$, where $i$ is the band index and $(k_1,k_2) \in \left(-\frac{\pi}{q}, \frac{\pi}{q}\right] \times (-\pi, \pi]$. To write explicitly, we have

$$\psi_{i,k_1,k_2}(x-q,y) = \mathrm{e}^{\mathrm{i}2\pi p y}\mathrm{e}^{-\mathrm{i}qk_1}\psi_{i,k_1,k_2}(x,y), \tag{S1.1}$$

$$\psi_{i,k_1,k_2}(x,y-1) = \mathrm{e}^{-\mathrm{i}k_2}\psi_{i,k_1,k_2}(x,y), \tag{S1.2}$$

which is the magnetic Bloch theorem.

Moreover, because $\widehat{M}(\boldsymbol{a}_1)$ also commutes with Hamiltonian, the states $\widehat{M}(r\boldsymbol{a}_1)\psi_{i,k_1,k_2} (r = 1, \dots, q-1)$ all have degenerate energy $\epsilon_{i\boldsymbol{k}}$. But these states hold different $k_2$ momentum of $k_2 - r2\pi\frac{p}{q} (\mathrm{mod}\ 2\pi)$ due to the fact that $\widehat{M}(\boldsymbol{a}_2)\left(\widehat{M}(r\boldsymbol{a}_1)\psi_{i,k_1,k_2}\right) = \mathrm{e}^{\mathrm{i}\left(r2\pi\frac{p}{q}\right)}\widehat{M}(r\boldsymbol{a}_1)\left(\widehat{M}(\boldsymbol{a}_2)\psi_{i,k_1,k_2}\right) = \mathrm{e}^{-\mathrm{i}\left(k_2 - r2\pi\frac{p}{q}\right)}\widehat{M}(r\boldsymbol{a}_1)\psi_{i,k_1,k_2}$. Thus, we find $\widehat{M}(r\boldsymbol{a}_1)\psi_{i,k_1,k_2}(\boldsymbol{r}) = \mathrm{e}^{\mathrm{i}\theta_{i,r}(\boldsymbol{k})}\psi_{i,k_1,k_2 - r2\pi\frac{p}{q}(\mathrm{mod}\ 2\pi)}(\boldsymbol{r})$ for some $\theta_{i,r}(\boldsymbol{k}) \in \mathbb{R}$. Or equivalently,

$$\psi_{i,k_1,k_2}(x-r,y) = \mathrm{e}^{\mathrm{i}\theta_{i,r}(k_1,k_2)}\mathrm{e}^{\mathrm{i}\left(r2\pi\frac{p}{q}y\right)}\psi_{i,k_1,k_2 - r2\pi\frac{p}{q}(\mathrm{mod}\ 2\pi)}(x,y), \tag{S2}$$

with a degenerate eigenenergy of $\epsilon_{i\boldsymbol{k}}$. This property (S2) is essential to our proof.

We prove the strong translational symmetry for electronic density at first. For $y$-direction, it's obvious that $\left|\psi_{i,k_1,k_2}(x,y-1)\right|^2 = \left|\psi_{i,k_1,k_2}(x,y)\right|^2$ by using Eq. (S1.2) and therefore $\rho(x,y-1) = \rho(x,y)$. For $x$-direction,

$$\begin{aligned}
\rho(x-1,y) &= \sum_i^{\mathrm{occ}} \sum_{(k_1,k_2)\in\left(-\frac{\pi}{q},\frac{\pi}{q}\right]\times(-\pi,\pi]} f_{i,k_1,k_2}\left|\psi_{i,k_1,k_2}(x-1,y)\right|^2 \\
&= \sum_i^{\mathrm{occ}} \sum_{(k_1,k_2)\in\left(-\frac{\pi}{q},\frac{\pi}{q}\right]\times(-\pi,\pi]} f_{i,k_1,k_2-2\pi\frac{p}{q}(\mathrm{mod}\ 2\pi)}\left|\psi_{i,k_1,k_2-2\pi\frac{p}{q}(\mathrm{mod}\ 2\pi)}(x,y)\right|^2 \\
&= \sum_i^{\mathrm{occ}} \sum_{(k_1,k_2)\in\left(-\frac{\pi}{q},\frac{\pi}{q}\right]\times(-\pi,\pi]} f_{i,k_1,k_2}\left|\psi_{i,k_1,k_2}(x,y)\right|^2 \\
&= \rho(x,y),
\end{aligned} \tag{S3}$$

where we use Eq. (S2) and the equation $f_{i,k_1,k_2} = f_{i,k_1,k_2-2\pi\frac{p}{q}(\mathrm{mod}\ 2\pi)}$ (due to degeneracy) in the second equation, and the fact $\sum_{k_2}\varphi(k_2) = \sum_{k_2}\varphi\left(k_2 - 2\pi\frac{p}{q}(\mathrm{mod}\ 2\pi)\right)$ for any function $\varphi$ in the third equation.

We subsequently prove the strong translational symmetry for current density. The proof for $y$-direction is straightforward and thus we only present the proof for $x$-direction,

$$\begin{aligned}\boldsymbol{j}(x-1,y) &= -\sum_{i}^{\text{occ}} \sum_{(k_1,k_2)\in\left(-\frac{\pi}{q},\frac{\pi}{q}\right]\times(-\pi,\pi]} f_{i,k_1,k_2} \Re\left[\psi_{i,k_1,k_2}^{*}(x-1,y)\left(-\mathrm{i}\boldsymbol{\nabla}+\boldsymbol{A}(x-1,y)\right)\psi_{i,k_1,k_2}(x-1,y)\right] \\ &= -\sum_{i}^{\text{occ}} \sum_{(k_1,k_2)\in\left(-\frac{\pi}{q},\frac{\pi}{q}\right]\times(-\pi,\pi]} f_{i,k_1,k_2} \Re\left[\psi_{i,k_1,k_2-2\pi\frac{p}{q}(\operatorname{mod} 2\pi)}^{*}(x,y)\mathrm{e}^{-\mathrm{i}\left(2\pi\frac{p}{q}y\right)}\left(-\mathrm{i}\boldsymbol{\nabla}+\boldsymbol{A}(x-1,y)\right)\mathrm{e}^{\mathrm{i}\left(2\pi\frac{p}{q}y\right)}\psi_{i,k_1,k_2-2\pi\frac{p}{q}(\operatorname{mod} 2\pi)}(x,y)\right] \\ &= -\sum_{i}^{\text{occ}} \sum_{(k_1,k_2)\in\left(-\frac{\pi}{q},\frac{\pi}{q}\right]\times(-\pi,\pi]} f_{i,k_1,k_2-2\pi\frac{p}{q}(\operatorname{mod} 2\pi)} \Re\left[\psi_{i,k_1,k_2-2\pi\frac{p}{q}(\operatorname{mod} 2\pi)}^{*}(x,y)\left(-\mathrm{i}\boldsymbol{\nabla}+\boldsymbol{A}(x,y)\right)\psi_{i,k_1,k_2-2\pi\frac{p}{q}(\operatorname{mod} 2\pi)}(x,y)\right] \\ &= -\sum_{i}^{\text{occ}} \sum_{(k_1,k_2)\in\left(-\frac{\pi}{q},\frac{\pi}{q}\right]\times(-\pi,\pi]} f_{i,k_1,k_2} \Re\left[\psi_{i,k_1,k_2}^{*}(x,y)\left(-\mathrm{i}\boldsymbol{\nabla}+\boldsymbol{A}(x,y)\right)\psi_{i,k_1,k_2}(x,y)\right] \\ &= \boldsymbol{j}(x,y), \end{aligned} \tag{S4}$$

where we use Eq. (S2) in the second equation and the relation $\mathrm{e}^{-\mathrm{i}\left(2\pi\frac{p}{q}y\right)}\left(-\mathrm{i}\boldsymbol{\nabla}+\boldsymbol{A}(x-1,y)\right)\mathrm{e}^{\mathrm{i}\left(2\pi\frac{p}{q}y\right)}\psi = \left[-\mathrm{i}\boldsymbol{\nabla}+\boldsymbol{A}(x-1,y)+2\pi\frac{p}{q}\hat{y}\right]\psi = [-\mathrm{i}\boldsymbol{\nabla}+\boldsymbol{A}(x,y)]\psi$ in the third equation. Hence, the proposition is proven.

The strong translational symmetry proved in this sector can be understood even more clearly through the arbitrariness of choices for the magnetic unit cell. Let us take the density as an example. If we expand the cell along $\boldsymbol{a}_1$ for a factor of $q$, we will get $\rho(\boldsymbol{r})=\rho(\boldsymbol{r}+q\boldsymbol{a}_1)=\rho(\boldsymbol{r}+\boldsymbol{a}_2)$; but if we expand the cell along $\boldsymbol{a}_2$ for a factor of $q$, we will get $\rho'(\boldsymbol{r})=\rho'(\boldsymbol{r}+\boldsymbol{a}_1)=\rho'(\boldsymbol{r}+q\boldsymbol{a}_2)$. Since choices for the magnetic unit cell should not change physical observables, the two densities must coincide $\rho(\boldsymbol{r})=\rho'(\boldsymbol{r})$, and therefore we have $\rho(\boldsymbol{r})=\rho(\boldsymbol{r}+\boldsymbol{a}_1)=\rho(\boldsymbol{r}+\boldsymbol{a}_2)$. The argument for current density is the same.

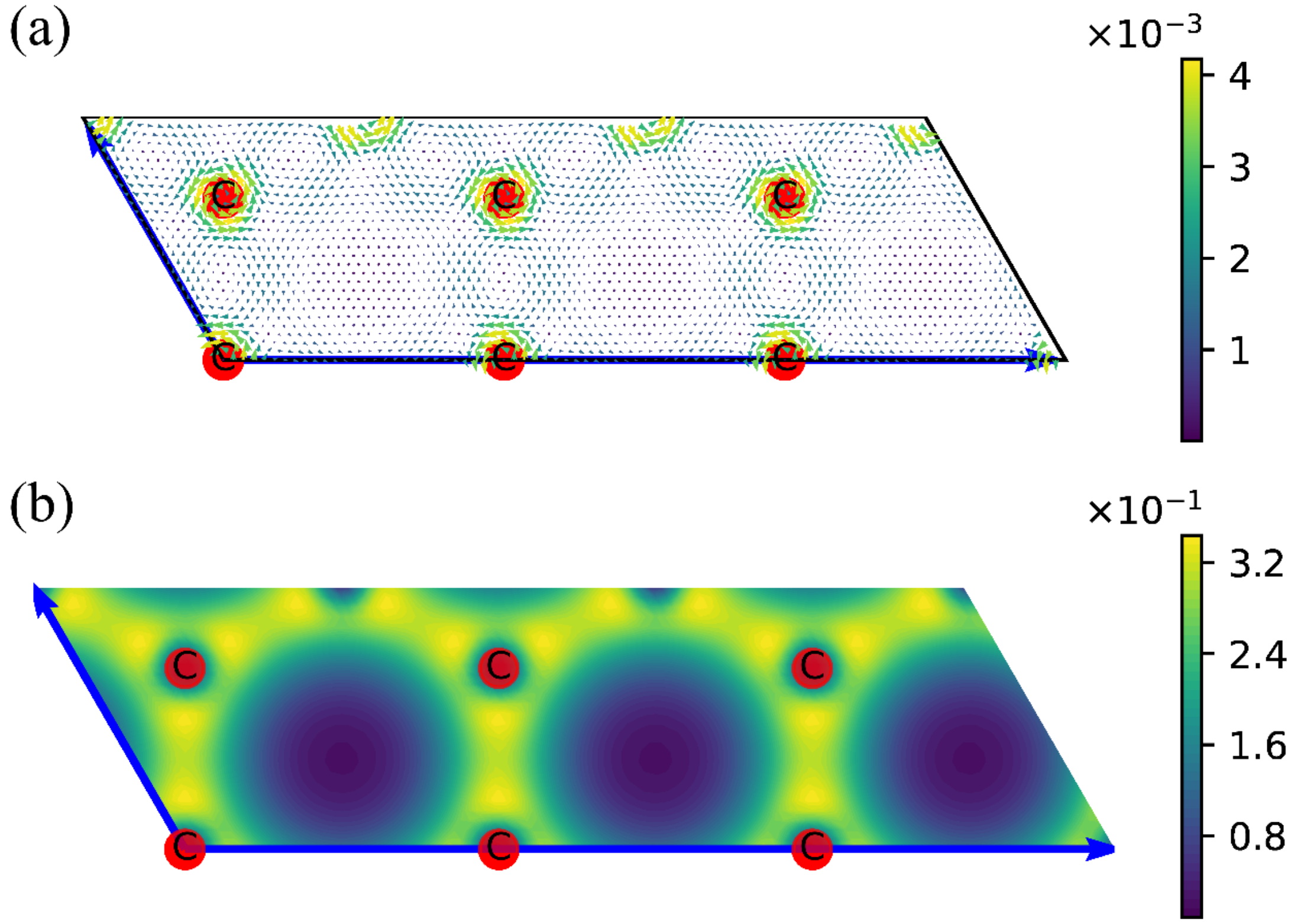

FIG. S1. Electronic density (a) and current density (b) of graphene under a magnetic flux of $n_\Phi = 2/3$ ($\approx$ 53440 T).

## 2. Magnetic band shift

In this sector, we would prove that the magnetic energy bands will suffer a shift within the Brillouin zone by the collective translation of atoms.

We denote the atomic translation as $\boldsymbol{d}$. The mechanical momentum remains the same after the translation $-\mathrm{i}\boldsymbol{\nabla}+\boldsymbol{A}(\boldsymbol{r})\rightarrow-\mathrm{i}\boldsymbol{\nabla}+\boldsymbol{A}(\boldsymbol{r})$ while the local potential and the nonlocal pseudopotential transform as $V_{\mathrm{per}}(\boldsymbol{r})\rightarrow V_{\mathrm{per}}(\boldsymbol{r}-\boldsymbol{d})$ and $\hat{V}_{\mathrm{GIPAW}}=\sum_{\boldsymbol{R},n}\mathrm{e}^{-\mathrm{i}\boldsymbol{A}(\boldsymbol{R})\cdot\boldsymbol{r}}|p_{\boldsymbol{R},n}\rangle a_n^{\boldsymbol{R}}\langle p_{\boldsymbol{R},n}|\mathrm{e}^{\mathrm{i}\boldsymbol{A}(\boldsymbol{R})\cdot\boldsymbol{r}}\rightarrow\sum_{\boldsymbol{R},n}\mathrm{e}^{-\mathrm{i}\boldsymbol{A}(\boldsymbol{R}+\boldsymbol{d})\cdot\boldsymbol{r}}|p_{\boldsymbol{R}+\boldsymbol{d},n}\rangle a_n^{\boldsymbol{R}}\langle p_{\boldsymbol{R}+\boldsymbol{d},n}|\mathrm{e}^{\mathrm{i}\boldsymbol{A}(\boldsymbol{R}+\boldsymbol{d})\cdot\boldsymbol{r}}$, respectively. To no surprise, we find all of three relations can be written as a unitary transformation $\hat{V}\rightarrow\hat{M}(\boldsymbol{d})\hat{V}\hat{M}^{\dagger}(\boldsymbol{d})$, where the magnetic translational operator is defined as $\hat{M}(\boldsymbol{d})=\mathrm{e}^{-\mathrm{i}\boldsymbol{A}(\boldsymbol{d})\cdot\boldsymbol{r}}\mathrm{e}^{-\mathrm{i}\boldsymbol{d}\cdot\hat{\boldsymbol{p}}}$. Thus, the Hamiltonian and eigenstates that correspond to the eigenvalue $\epsilon_{i\boldsymbol{k}}$ should transform as

$$\hat{H}\rightarrow\hat{H}^{\boldsymbol{d}}\coloneqq\hat{M}(\boldsymbol{d})\hat{H}\hat{M}^{\dagger}(\boldsymbol{d}),\tag{S5}$$

and

$$\psi_{i\boldsymbol{k}}(\boldsymbol{r})\rightarrow\psi_{i\boldsymbol{k}}^{\boldsymbol{d}}(\boldsymbol{r})\coloneqq\hat{M}(\boldsymbol{d})\psi_{i\boldsymbol{k}}(\boldsymbol{r})=\mathrm{e}^{-\mathrm{i}\boldsymbol{A}(\boldsymbol{d})\cdot\boldsymbol{r}}\psi_{i\boldsymbol{k}}(\boldsymbol{r}-\boldsymbol{d}),\tag{S6}$$

respectively.

However, the transformed eigenstate $\psi_{i\boldsymbol{k}}^{\boldsymbol{d}}$ has a different $\boldsymbol{k}$ momentum from $\psi_{i\boldsymbol{k}}$. After some simple derivation, we find

$$\psi_{i\boldsymbol{k}}^{\boldsymbol{d}}(\boldsymbol{r}+\boldsymbol{T})=\mathrm{e}^{\mathrm{i}(\boldsymbol{k}-\boldsymbol{q}(\boldsymbol{d}))\cdot\boldsymbol{T}}e^{-\mathrm{i}\boldsymbol{A}(\boldsymbol{T})\cdot\boldsymbol{r}}\psi_{i\boldsymbol{k}}^{\boldsymbol{d}}(\boldsymbol{r}),\tag{S7}$$

where $\boldsymbol{T}$ denotes the translational vector that is an integral linear combination of cell basis and $\boldsymbol{q}(\boldsymbol{d})$ satisfies $\boldsymbol{q}(\boldsymbol{d})\cdot\boldsymbol{T}=\boldsymbol{A}(\boldsymbol{d})\cdot\boldsymbol{T}-\boldsymbol{A}(\boldsymbol{T})\cdot\boldsymbol{d}$. Consequently, $\psi_{i\boldsymbol{k}}^{\boldsymbol{d}}$, which corresponds to the eigenvalue $\epsilon_{i\boldsymbol{k}}$, has the $\boldsymbol{k}$-momentum of $\boldsymbol{k}-\boldsymbol{q}(\boldsymbol{d})$ (mod $2\pi\boldsymbol{b}^{\alpha},\alpha=1,2,3$). That is, the magnetic bands are shifted by a translation-dependent vector of $-\boldsymbol{q}(\boldsymbol{d})$, which concludes the proof. For Landau gauge $\boldsymbol{A}_{\mathrm{Lan}}(\boldsymbol{r})\coloneqq2\pi n_{\Phi}(\boldsymbol{b}^1\cdot\boldsymbol{r})\boldsymbol{b}^2$, $\boldsymbol{q}(\boldsymbol{d})$ is chosen as

$$\boldsymbol{q}(\boldsymbol{d})=2\pi n_{\Phi}[(\boldsymbol{b}^1\cdot\boldsymbol{d})\boldsymbol{b}^2-(\boldsymbol{b}^2\cdot\boldsymbol{d})\boldsymbol{b}^1].\tag{S8}$$

This phenomenon can also be explained by the non-commutativity between magnetic translational operators: since $\hat{M}(\boldsymbol{d})$ doesn't commute with $\hat{M}(\boldsymbol{a}_1)$ and $\hat{M}(\boldsymbol{a}_2)$ in general, $\hat{M}(\boldsymbol{d})\psi_{i\boldsymbol{k}}$ cannot have the same eigenvalues of these operators as $\psi_{i\boldsymbol{k}}$.

We give a specific example of silicon about "magnetic band shift" in this paragraph. We choose $\boldsymbol{d}=\frac{1}{2}\boldsymbol{a}_1$, and compute the total energy, eigenvalues and density of states (DOS) before and after the atomic translation for $n_{\Phi}=1,2$ ($\approx$ 40141,80283 T) which correspond to $\boldsymbol{q}(\boldsymbol{d})=\pi\boldsymbol{b}^2,2\pi\boldsymbol{b}^2$, respectively. Therefore, to preserve the band structure following the atomic translation $\boldsymbol{d}$, the second fractional coordinate of the $\boldsymbol{k}$-points must be adjusted by subtracting 0.5 (mod 1) for $n_\Phi = 1$, while keeping the same $\boldsymbol{k}$-points for $n_\Phi = 2$. The total energy and eigenvalues for some representative $\boldsymbol{k}$-points of the lowest five bands are listed in Table S1. The DOS are plotted in Fig. S2., from which we can observe silicon suffer a phase transition from insulator to metal when applying the magnetic field [2]. The computation is performed on a $8 \times 8 \times 8$ $\boldsymbol{k}$-point grid at an energy cutoff of 20 Hartree.

Table S1. Total energy and eigenvalues for some representative $\boldsymbol{k}$-points of the lowest five bands of silicon under a magnetic field before and after the atomic translation.

| | $n_\Phi = 1$ | | $n_\Phi = 2$ | |
|---|---|---|---|---|
| | Energy before translation (Ha) | Energy after a translation of $0.5\,\boldsymbol{a}_1$ (Ha) | Energy before translation (Ha) | Energy after a translation of $0.5\,\boldsymbol{a}_1$ (Ha) |
| Total Energy (Ha) | -8.4173675 | -8.4173676 | -8.3750565 | -8.3750568 |
| $\boldsymbol{k}$-point | (-0.375, 0.125, -0.375) | (-0.375, -0.375, -0.375) | (-0.375, 0.125, -0.375) | (-0.375, 0.125, -0.375) |
| 1 | -0.115364 | -0.115364 | -0.059046 | -0.059046 |
| 2 | 0.016564 | 0.016564 | 0.005221 | 0.005221 |
| 3 | 0.115942 | 0.1159412 | 0.079297 | 0.079297 |
| 4 | 0.1775285 | 0.1775285 | 0.186484 | 0.186485 |
| 5 | 0.263474 | 0.263474 | 0.304088 | 0.304088 |
| $\boldsymbol{k}$-point | (-0.375, -0.125, -0.250) | (-0.375, 0.375, -0.250) | (-0.375, -0.125, -0.250) | (-0.375, -0.125, -0.250) |
| 1 | -0.142624 | -0.142624 | -0.076681 | -0.076681 |
| 2 | 0.066963 | 0.066963 | -0.002892 | -0.002891 |
| 3 | 0.128379 | 0.128379 | 0.131996 | 0.131996 |
| 4 | 0.203690 | 0.203690 | 0.183773 | 0.183773 |
| 5 | 0.278290 | 0.278292 | 0.300507 | 0.300307 |
| ***k*-point** | (0.000, 0.000, 0.000) | (0.000, 0.500, 0.000) | (0.000, 0.000, 0.000) | (0.000, 0.000, 0.000) |
| 1 | -0.158880 | -0.158880 | -0.091728 | -0.091729 |
| 2 | 0.068347 | 0.068347 | -0.012581 | -0.012581 |
| 3 | 0.211877 | 0.211877 | 0.163453 | 0.163453 |
| 4 | 0.214504 | 0.214504 | 0.221860 | 0.221860 |
| 5 | 0.252458 | 0.252458 | 0.313947 | 0.313947 |

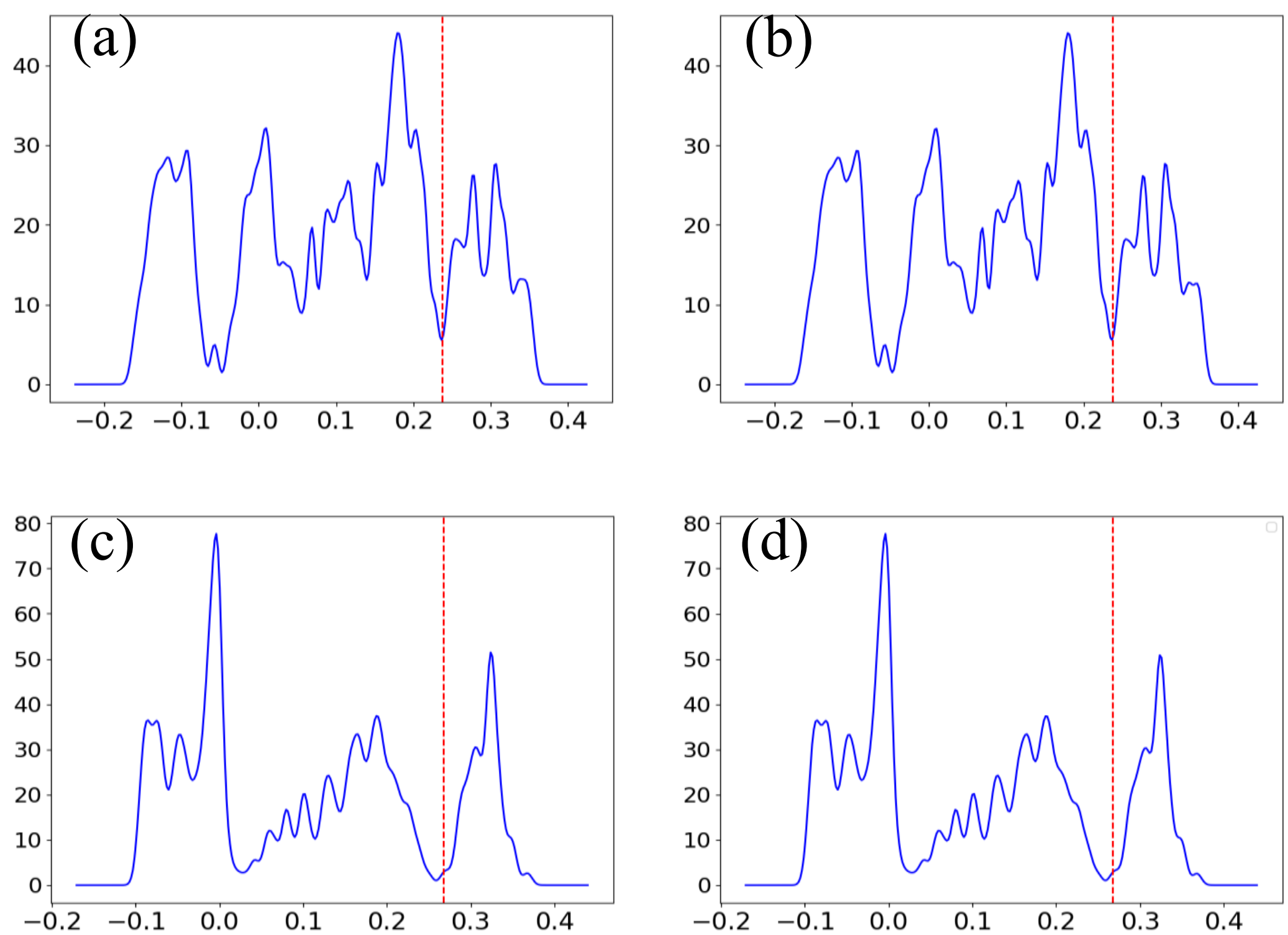

FIG. S2. Density of states of silicon under a magnetic field before and after the atomic translation of $\frac{1}{2}\boldsymbol{a}_1$, where the red dotted line denotes the Fermi energy. (a) $n_\Phi = 1\ (\approx 40141\ \text{T})$ before the translation. (b) $n_\Phi = 1\ (\approx 40141\ \text{T})$ after the translation. (c) $n_\Phi = 2\ (\approx 80283\ \text{T})$ before the translation. (d) $n_\Phi = 2\ (\approx 80283\ \text{T})$ after the translation.

## 3. Details of the calculations

**Test on molecules**

For molecular magnetizabilities, DFT calculations are performed at the generalized gradient approximation (GGA) level using PBE functional [3] with the energy cut-off of $40$ Hartree in a cubic supercell of size $12\text{Å} \times 12\text{Å} \times 12\text{Å}$, and the convergence criteria of $10^{-6}$ Hartree. We have also ensured using larger supercells (i.e., smaller magnetic field) gives similar results by our tests and therefore our results are free of non-linear effects caused by the external field, as shown in Table S2 where the molecules are computed in three supercells of different sizes.

Table S2. Molecular magnetizabilities in units of $10^{-30}$ J $\text{T}^{-2}$ calculated via three supercells of different sizes. The magnetic field strengths corresponding to each supercell are 2871.4 T, 2109.6 T and 1615.1 T, respectively.

| | cell (12Å×12Å×12Å) | cell (14Å×14Å×12Å) | cell (16Å×16Å×12Å) |
|---|---|---|---|
| Be | -227.0 | -226.5 | -226.5 |
| $H_2$ | -71.2 | -71.5 | -71.1 |
| CO | -299.2 | -296.9 | -300.6 |
| HF | -175.4 | -174.4 | -177.0 |
| $H_2O$ | -239.5 | -239.3 | -240.2 |
| AlH | 269.5 | 272.3 | 272.2 |
| BH | 504.6 | 504.3 | 504.6 |

For the calculations of benzene, the calculation is performed with energy cut-off of 20 Hartree in a cuboid supercell of size $12\text{Å} \times 12\text{Å} \times 9\text{Å}$ with $n_\Phi = 1$.

**Test on solids**

For solid-state magnetizabilities, DFT calculations are performed within the local density approximation (LDA) [4]. For argon and krypton solids at FCC phases, calculations are carried out on a $6 \times 6 \times 6$ $\boldsymbol{k}$-point grid with the energy-cutoff of 40 Hartree. For semiconductors, namely diamond ($C_2$) and silicon ($Si_2$), we expand the conventional FCC cell along the direction of $\boldsymbol{a}_1$ by a factor of three to reduce the magnetic strength, and then perform calculations on a $2 \times 6 \times 6$ $\boldsymbol{k}$-point grid with energy-cutoffs of 30 and 20 Hartree, respectively. The convergence criteria are $10^{-6}$ Hartree. We also carry out the supercell test for diamond in Table S3.

Table S3. The magnetizability of diamond in units of $10^{-6}\,\mathrm{cm}^3/\mathrm{mole}$ calculated via two supercells with sizes three times and four times of the unit cell respectively. The magnetic field strengths corresponding to each supercell are 10870.4 T and 8152.8 T, respectively.

| | cell (3×1×1) | cell (4×1×1) |
|---|---|---|
| diamond | -11.16 | -10.96 |

For calculating the band structures of graphene, we use the PBE functional with energy cutoffs of 40 Hartree. And for the Berry curvatures and Chern numbers of graphene, we use the PBE functional with the energy cutoff of 30 Hartree and a $\boldsymbol{k}$-point grid of size $40 \times 40 \times 1$.

## 4. Density variation of benzene under the magnetic field

For all diamagnetic molecules we have computed, we observe an increase not only in the total energy but also in the absolute values of both kinetic energy and local potential energy, indicating the induced precession and localization of electrons by the external magnetic field [5,6]. To visualize this phenomenon, we plot the density variation of benzene after and before the application of the field, as shown in Fig. S3. For (closed-shell) paramagnetic molecules, in contrast, both the total energy and the absolute values of kinetic energy and local potential energy decrease when subjected to a magnetic field.

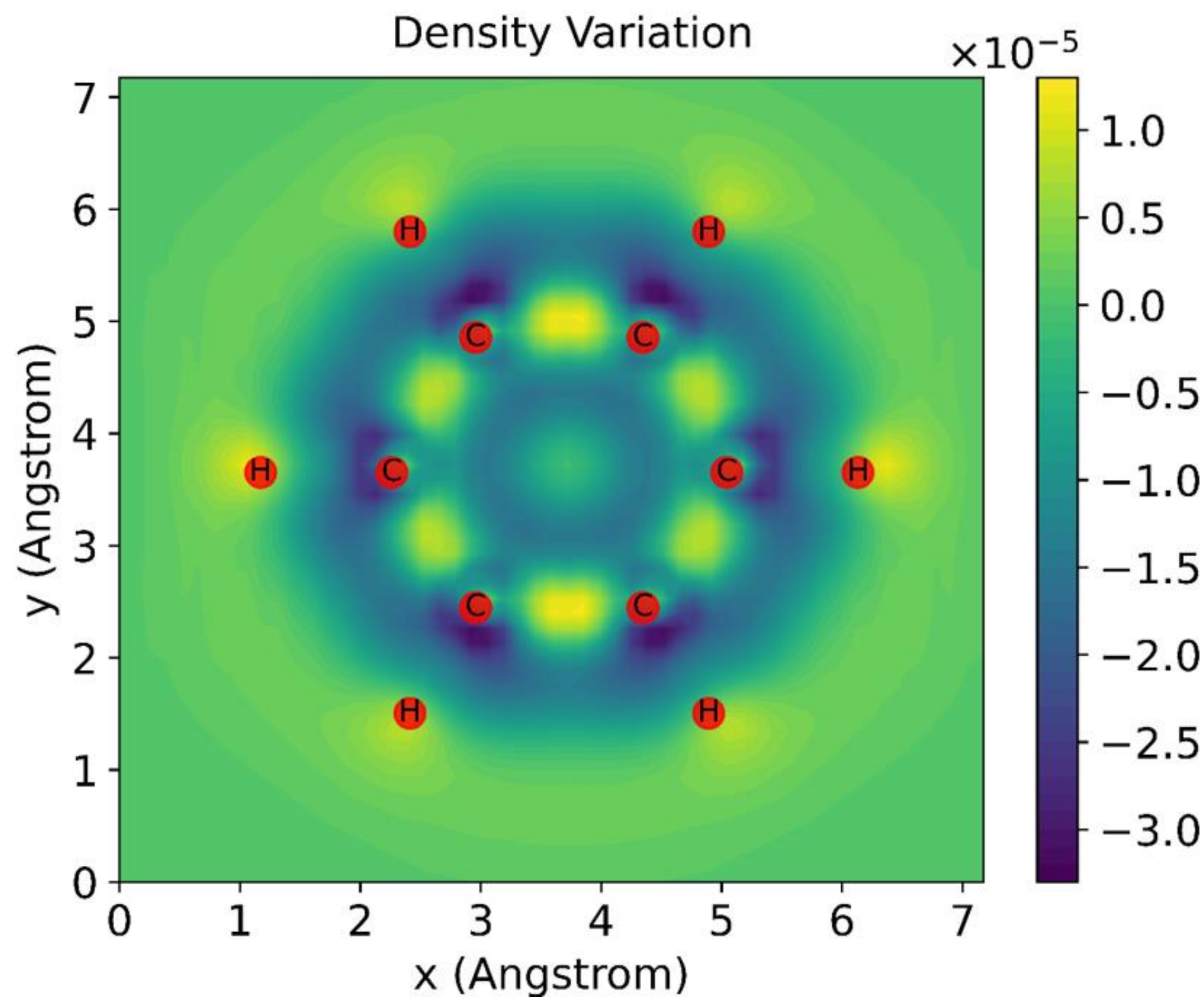

FIG. S3 Density Variation of benzene under the magnetic field of approximately 2871 T.